\documentclass[12pt,preprint]{aastex}
\usepackage{graphicx}
\usepackage{ifpdf}
\newcommand{\met}{\hbox{E\kern-0.5em\lower-0.1ex\hbox{/}}_T}
\newcommand\simlt{\lower.5ex\hbox{$\; \buildrel < \over \sim \;$}}
\newcommand\simgt{\lower.5ex\hbox{$\; \buildrel > \over \sim \;$}}

\begin{document}
\title{Interaction of a magnetized shell with an ambient medium: limits on impulsive magnetic acceleration}
\author{Amir Levinson}
\altaffiltext{1}{School of Physics \& Astronomy, Tel Aviv University,
Tel Aviv 69978, Israel; Levinson@wise.tau.ac.il}

\begin{abstract}
The interaction of relativistic magnetized ejecta with an ambient medium is studied for a range of structures and 
magnetization of the unshocked ejecta.  We particularly focus on the effect of the ambient medium on the dynamics of 
an impulsive, high-sigma shell.  It is found 
that for sufficiently high values of the initial magnetization $\sigma_0$ the evolution of the system is  significantly 
altered by the ambient medium  well before the shell reaches its coasting phase.  The maximum Lorentz factor of
the shell is limited to values well below  $\sigma_0$; for a shell of initial energy $E=10^{52}E_{52}$ erg
and size $r_0=10^{12}T_{30}$ cm expelled into a medium having a uniform density $n_i$ we obtain 
$\Gamma_{\rm max}\simeq180(E_{52}/T_{30}^3 n_i)^{1/8}$ in the high sigma limit.   The reverse shock and any internal shocks
that might form if the source is fluctuating are shown to be very weak.   The restriction on the Lorentz factor is more severe 
for shells propagating in a stellar wind.  Intermittent ejection of small sub-shells doesn't seem to help, as the shells merge 
while still highly magnetized.
Lower sigma shells start decelerating after reaching the coasting phase and spreading away.   
The properties of the reverse shock then depend on the density profiles of the coasting shell and the ambient medium.
For a self-similar cold shell the reverse shock becomes strong as it propagates inwards, and the system  
eventually approaches the self-similar solution derived recently by Nakamura \& Shigeyama.    
\end{abstract}
\section{Introduction}
The interaction of relativistic ejecta with the surrounding medium is an issue of considerable interest.  
During the early stages of the evolution a double shock structure forms, consisting of a forward shock that propagates
in the ambient medium, a reverse shock crossing the ejecta and a contact interface separating
the shocked ejecta and the shocked ambient medium.  Under certain conditions, an observable flash of electromagnetic
radiation is expected to be emitted during the propagation of the reverse shock.  In the fireball scenario commonly adopted,
the naive expectation has been that optical flashes associated with reverse shock crossing should be quite common (e.g., Kumar \& Panaitescu
2003; but c.f. Nakar \& Piran 2004), however,   
despite considerable observational efforts only a few have been detected, indicating that such flashes are rare.

It has been proposed that the paucity of optical flashes may be attributed to an early onset of a R-T instability (Levinson, 2010a,b),
or strong magnetization of the ejecta (e.g., Zhang \& Kobayashi 2005; Mimica et al. 2009).   The latter is anticipated if the free
energy is extracted magnetically in the form of a Poynting-flux- dominated flow (e.g., Levinson \& Eichler 1993; Lyutikov \& Blandford 2003
; Giannios \& Spruit 2005).  Stationary magnetic outflows allow, in general, 
only partial conversion of magnetic-to-kinetic energy, implying high magnetization at the onset of the afterglow phase.  A better conversion
can be achieved if the outflow is collimated into a small opening angle, $\theta\simeq\Gamma^{-1}$ (Komissarov et al. 2009), though 
corking from a star, as in the collapsor model for GRBs, can alleviate the latter condition (Tchekhovskoy et al., 2010; Komissarov et al. 2010).
However, even then $\sigma\sim1$ is anticipated at best (Lyubarski 2009). 

Recently it has been shown (Granot et al., 2010; hereafter GKS10; Lyutikov 2010a,b) that time-dependent effects 
may play a crucial role in the acceleration 
of a magnetized flow.   These authors considered the acceleration of a spherical, impulsive high sigma shell of initial width $\Delta=r_0$ and 
magnetization $\sigma_0$, expelled by a central source.  They have shown that, unlike a stationary flow for which acceleration
ceases at $\Gamma_\infty\sim\sigma_0^{1/3}$, $\sigma_\infty\sim\sigma_0^{2/3}$ , 
the impulsive shell continues accelerating even after loosing causal contact 
with the central source until reaching nearly complete conversion of magnetic energy into bulk kinetic energy. 
The terminal Lorentz factor of the shell is $\Gamma_\infty\simeq \sigma_0$.  During the acceleration phase, that they term 
``magnetic rocket acceleration'', the major fraction of the 
shell energy is contained in a layer of width $2r_0$, bounded between the front of a rarefaction wave reflected from the central source and the 
head of the shell.   The average Lorentz factor of the shell, roughly equals the Lorentz factor of the fluid at the rarefaction front, evolves as 
$<\Gamma>\propto t^{1/3}$.  The structure of this layer is well described by a self-similar solution.  Once the shell enters the coasting phase 
its width starts growing and its magnetization continues to drop.  

 In this paper we consider the interaction of the shell with the external medium
 and show that in the high-sigma limit the evolution of the system is dramatically altered.    A preliminary account of 
 the effect of the ambient medium is given in GKS10, and a more detailed discussion in Lyutikov (2010a,b).
Specifically, it is shown that for initial magnetization $\sigma_0$ larger than some critical value, deceleration of 
the contact interface commences well before the shell has reached the coasting phase, when it is still highly magnetized.  
The Lorentz factor of the contact discontinuity evolves as $\Gamma_c\propto t^{-1/2}$, while the rear boundary of the unshocked
shell is still accelerating.  The maximum Lorentz factor of the compressed shell is then limited to $\Gamma<<\sigma_0$.  The reverse shock in this 
case is very weak or nonexistent, and within the framework of ideal MHD no internal dissipation is practically expected (by either the 
reverse shock or any internal shocks that might form in a multi-shell scenario) before the onset of the afterglow phase.
Lower sigma shells start decelerating after reaching the coasting phase.   The properties of the reverse shock then depend on the 
density profiles of the coasting shell and the ambient medium, as discussed in \S \ref{sec:mag-shell}.  

\begin{figure}[h]
\centering
\includegraphics[width=11.0cm]{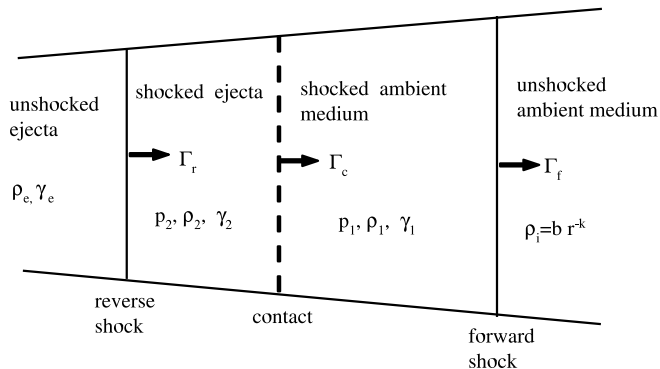}
\caption{\label{fig:a}Schematic representation of the double-shock system.  There are three characteristic surfaces:
a forward shock propagating in the ambient medium, a reverse shock sweeping the ejecta, and a contact discontinuity separating the
shocked ejecta and the shocked ambient medium.  The Lorentz factors of the three surfaces, measured with respect to the unshocked 
ambient medium, are indicated.  Quantities in the shocked ambient medium and shocked ejecta are denoted by subscripts 1 and 2, respectively.}
\end{figure}
\section{\label{sec:un_contact}Thin Shell Model}
We consider the interaction of a cold, magnetized shell with an ambient medium having a density profile 
$\rho_i(r)=a_ir^{-k}$.  The ejecta is characterized by a Lorentz factor $\gamma_e(r,t)$, 
density $\rho_e(r,t)$ and magnetic field vector $b^\mu_e(r,t)$, assumed to be given, where the 4-vector of the magnetic
field, $b^\mu$, is defined in appendix \ref{sec:app-jump-r}.  To simplify the analysis we shall assume a spherical 
shell with a purely toroidal magnetic field, viz., $b_e^\mu=(0,0,0,b_e)$.  The magnetic pressure of the shell is then $b_e^2/2$
and the corresponding sigma parameter is $\sigma_e=b^2_e/\rho_e$. 
The structure of the shocked shell is shown schematically in Fig. \ref{fig:a}.  The subscript 1 refers to 
the shocked ambient medium and 2 to the shocked shell.  The Lorentz factors of the
forward shock, reverse shock and the contact discontinuity are denoted by $\Gamma_{f}(t)$,
$\Gamma_{r}(t)$ and $\Gamma_c(t)$, respectively, and satisfy the relation $\Gamma_{r}< \Gamma_{c}<\Gamma_{f}$.

The thin shell approximation assumes that the shocked layers are uniform.  Then $\gamma_1=\gamma_2=\Gamma_c$.
For the situations envisaged here the forward shock can be considered ultra-relativistic, $\Gamma_f>>1$.  The jump
conditions at the forward shock then yield $\Gamma_f=\sqrt{2}\Gamma_c$, and
\begin{equation}
p_1=\frac{4}{3}\rho_{i}\Gamma_c^2\label{jump-forward}.
\end{equation}

The reverse shock, on the other hand, cannot be considered ultra-relativistic in general and, therefore, a 
complete treatment is required.   The jump conditions at the reverse shock, derived in appendix \ref{sec:app-jump-r}, 
yield the relations
\begin{eqnarray}
\quad  \rho_{2}=\rho_e h(q_2,q_e),\quad, b_2=b_e h(q_2,q_e),\quad p_{2}=\rho_ef(q_2,q_e,\sigma_e),\label{jump-rev}
\end{eqnarray}
subject to the condition
\begin{equation}
\Psi(q_2,q_e,\sigma_e)=0\label{Psi0},
\end{equation}
here $q_2=(\Gamma_c/\Gamma_r)^2$, $q_e=(\gamma_e/\Gamma_r)^2$, and the functions $f$, $h$ and $\Psi$ are defined in 
Eqs.  (\ref{hq})-(\ref{psiq}).  Pressure balance at the contact, viz., $p_1=p_2+b_2^2/2$, yields 
$(4/3)\rho_i\Gamma_c^2=\rho_ef+b_e^2h^2/2=\rho_e(f+\sigma_eh^2/2)$, where Eqs. (\ref{jump-forward}) and (\ref{jump-rev})
have been employed.  Dividing the latter equation by $(4/3)\rho_i\gamma_e^2$, noting 
that $\Gamma_c^2/\gamma_e^2=q_2/q_e$, and defining
\begin{equation}
G(r,t)=3\rho_e/(4\rho_i\gamma_e^2),
\end{equation}
one finally arrives at
\begin{equation}
q_2/q_e=G(R_r,t)[f(q_2,q_e,\sigma_e)+\sigma_e(R_r,t)h^2(q_2,q_e)/2].\label{contac-p}
\end{equation}
The functions $G(r,t)$ and $\sigma_e(r,t)$ in Eq. (\ref{contac-p}) are computed just upstream of the reverse shock, at $r=R_r(t)$, 
where 
\begin{equation}
R_{r}(t)=R_{r0}+\int_{t_0}^t V_rdt^\prime=R_{r0}+\int_{t_0}^t{\left(1-\frac{1}{2\Gamma_r^2}\right)dt^\prime} \label{r-trajectory}
\end{equation}
is the trajectory of the reverse shock, given to order $O(\Gamma_r^{-2})$ by approximating the velocity of the reverse shock 
as $V_r\simeq 1-1/2\Gamma_r^2$.  Here $t=t_0$ is the initial time of impact.   Equations (\ref{Psi0}), (\ref{contac-p}), 
and (\ref{r-trajectory}) determine the evolution of the variables $q_2(t)$, $q_e(t)$, and $R_r(t)$ once $G(r,t)$ and $\sigma_e(r,t)$
are specified.

The shock compression ratio is given by
\begin{equation}
r=\frac{\rho_2\gamma_2^\prime}{\rho_e\gamma_e^\prime}=\frac{(q_2+1)(q_e-1)}{(q_2-1)(q_e+1)},\label{sh-compression}
\end{equation}
where $\gamma_e^\prime=\gamma_e\Gamma_r(1-v_eV_r)$ and $\gamma_2^\prime=\gamma_2\Gamma_r(1-v_2V_r)$ are the Lorentz factors
of the unshocked and shocked ejecta, as measured in the shock frame.  For an unmagnetized, relativistic shock $q_e>>1$, $q_2\simeq 2$
and $r=3$ as it should.    The fast magnetosonic Mach number of the flow just upstream the reverse shock is given by
\begin{equation}
M_A(t)=\frac{u_e^\prime}{\sqrt{\sigma_e(R_r,t)}}=\frac{q_e-1}{2\sqrt{q_e}\sqrt{\sigma_e(R_r,t)}},\label{Alfven-M}
\end{equation}
with $u_e^\prime=(\gamma_e^{\prime 2}-1)^{1/2}$.  The reverse shock exists as long as $M_A>1$.  For $M_A<1$ the compression 
of the shell is communicated by a magnetosonic wave that propagates from the contact discontinuity backwards in the fluid rest frame.

\subsection{\label{sec:unmag}Unmagnetized shell}
The limit of very low magnetization simplifies to $\sigma_e=0$ in the above equations.   
To illustrate the properties of the solutions we examine two situations.  The first one is that of 
a uniform shell, $\partial_r\gamma_e=\partial_r\rho_e=0$.    For a non-expanding shell mass conservation implies 
$\rho_e=\rho_{e0}(t/t_0)^{-2}$ and $G(R_r,t)=(3\rho_{e0}/4a_i)\gamma_e^{-2}(t/t_0)^{k-2}$.    For sufficiently small 
values of $G$ the reverse shock is relativistic, yielding $\Gamma_c^2\simeq2\Gamma_r^2$, $\gamma_e^2>>\Gamma_r^2$,
or equivalently $q_2\simeq2$, $q_e>>1$.  Eq. (\ref{fq}) with $\sigma_e=0$, $a=4$, $q_2=2$, $q_e>>1$ gives $f(q_2,q_e)\simeq q_e/6$.
Substituting the above results into Eq. (\ref{contac-p}) one finally obtains
\begin{equation}
\Gamma_c(t)=\gamma_e^{1/2}(\rho_{e0}/4a_i)^{1/4}(t/t_0)^{(k-2)/4},
\end{equation}
recovering earlier results (Sari \& Piran 1995).   

\begin{figure}[ht]
\centering
\includegraphics[width=12.0cm]{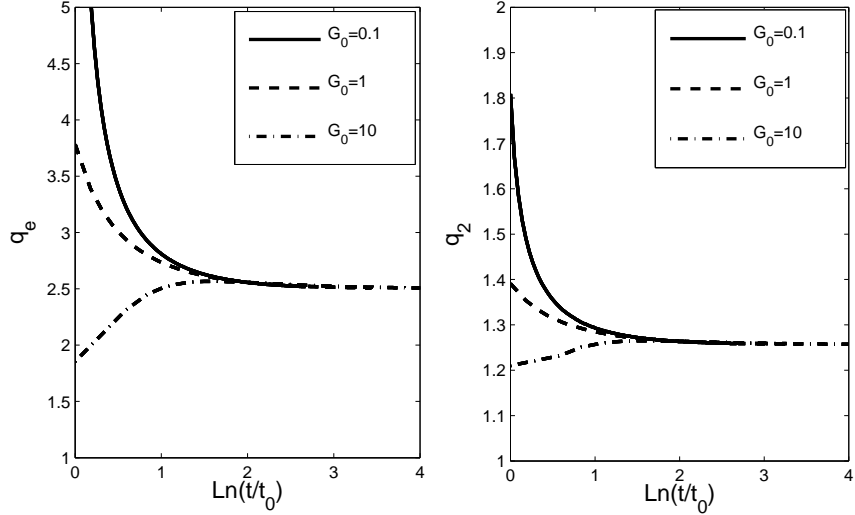}
\caption{\label{fig:un_3} The evolution of $q_e=(\gamma_e/\Gamma_r)^2$ (left panel) and $q_2=(\Gamma_c/\Gamma_r)^2$ (right panel)
for adiabatic index $\hat{\gamma}=5/3$, $k=0$ (a constant density medium), $n=2$, and different values of the parameter 
$G_0\equiv 3a_e/4a_i\Gamma_0^2$. }
\end{figure}

Our second example is the self similar ejecta considered by  Nakamura and Shigeyama (2006; NS06), and
discussed in appendix \ref{sec:app-self-eject}.  In this case  $\gamma_e(\chi)=\Gamma_0\chi^{-1/2}$ and the density profile is unrestricted.
Here $\chi=(1+2\Gamma_0^2)(1-r/t)$ is the self-similar parameter of the unshocked shell and $\Gamma_0$ is the initial 
Lorentz factor at the shell's head ($\chi=1$) just before impact (at $t=t_0$). 
The location of the reverse shock is given by
\begin{equation}
\chi_{r}(t)=(1+2\Gamma_0^2)(1-R_r/t).
\end{equation}
Differentiating the latter equation, 
using $dR_r/dt=1-1/2\Gamma_r^2$ from Eq. (\ref{r-trajectory}), and omitting terms of order $\Gamma_0^{-2}$ and higher, gives
\begin{equation}
\frac{d}{dt}(t\chi_{r})=(\Gamma_0/\Gamma_r)^2=\frac{q_e(\chi_r)\Gamma_0^2}{\gamma^2_e(\chi_r,t)}.\label{dchi}
\end{equation}
The initial conditions at $t=t_0$ are $\chi_r=1$, $\gamma_e(\chi_r)=\Gamma_0$.   
For illustration we adopt  $\rho_e=a_e(t/t_0)^{-3}\chi^{n/2}$ for which
 \begin{equation}
G(\chi_r,t)=(3a_e/4a_i\Gamma_0^2)(t/t_0)^{k-3}\chi_r^{1+n/2}.\label{G-self}
\end{equation}
Equations (\ref{dchi}), (\ref{Psi0}) and (\ref{contac-p}), with $\sigma_e=0$ and $G$ given by (\ref{G-self}),  are solved
simultaneously to yield $\Gamma_c(t)$, $\Gamma_r(t)$ and $\chi_r(t)$.

A particular solution can be sought for which  $q_e$, $q_2$ are constants.
Equation (\ref{dchi}) with  $\gamma_e(\chi_r)/\Gamma_0=\chi_r^{-1/2}$ readily yields
\begin{equation}
\chi_r(t)=(t/t_0)^{(q_e-1)}, \quad \Gamma_r=q^{-1}_e\Gamma_0(t/t_0)^{(1-q_e)/2},\quad \Gamma_c=\sqrt{q_e/q_2}\Gamma_r.\label{sol_self_2}
\end{equation}
Eq. (\ref{contac-p})  with  $dq_e/dt=dq_2/dt=0$ and $\sigma_e=0$ implies $dG/dt=0$, and using (\ref{G-self})  
one finds
\begin{equation}
q_e-1=\frac{6-2k}{n+2},\label{qe-sol-self}
\end{equation}
in accord with the self-similar solution derived by NS06.   The value of $q_2$ is computed numerically
from Eq. (\ref{Psi0}).    For any initial condition
different than the value given in Eq. (\ref{qe-sol-self}) self-similarity is broken and $q_e$, $q_2$ must evolve with time.
However, it is found that for the range of values for which the NS06 solution is applicable the dynamics of the shell eventually approaches
the self-similar limit, as naively expected since there is no scale in the problem.  This is demonstrated in Fig \ref{fig:un_3}, where solutions for $q_e$ and $q_2$ are plotted for different initial conditions.  As seen, the evolution quickly converges to the self-similar solution, whereby $q_e=2.5$ and $q_2=1.26$ for the choice of parameters in this example ($k=0$, $n=2$).

\subsection{\label{sec:mag-shell}Magnetized Shell}
The structure of an impulsive high-$\sigma$ shell has been computed recently by GKS10 and Lyutikov (2010a,b).  They
considered a situation in which a shell, initially at rest, is expanding into vacuum by pushing against a conducting wall.
The shell is assumed to be uniform initially with a density $\rho_0$, magnetization $\sigma_0=b_0^2/\rho_0>>1$, and finite width $r_0$. 
For a spherical shell the total energy is $E=(4\pi/3)r_0^3\rho_0\sigma_0$.
The dynamics of the shell follows several phases.  At  $t=0$ the shell starts accelerating, a simple rarefaction wave forms and propagates from the the shell's head towards the wall.  At $t=r_0$ the rarefaction wave reaches the wall, reflects, and starts propagating back  towards the head of the shell.   At this point the shell loses causal contact with the wall.  The major fraction of the energy (as measured in the Lab frame) is  contained in a layer of thickness $\sim2r_0$ bounded by the reflected  wave (at the rear end) and the shell's head (see appendix \ref{sec:app_impulsiv}). The structure of this layer can be described by a self similar solution.   The rear boundary of the shell accelerates as  $\gamma_w\simeq\sqrt{2}\sigma_0^{2/3}(t/r_0)^{1/6}$ (corresponding to the local magnetosonic speed in the fluid frame) and the fluid at the boundary as $\Gamma_\star\simeq(\sigma_0t/2r_0)^{1/3}$.  The head moves at a maximum Lorentz factor $\Gamma_0=2\sigma_0$.
In terms of the self-similar variable 
\begin{equation}
\chi=8\sigma^2_0[1-(r-2r_0)/t],\label{chi_impuls}
\end{equation}
the solution for the magnetization, density and Lorentz factor is given approximately by (see appendix \ref{sec:app-self-eject})
\begin{figure}[ht]
\centering
\includegraphics[width=12cm]{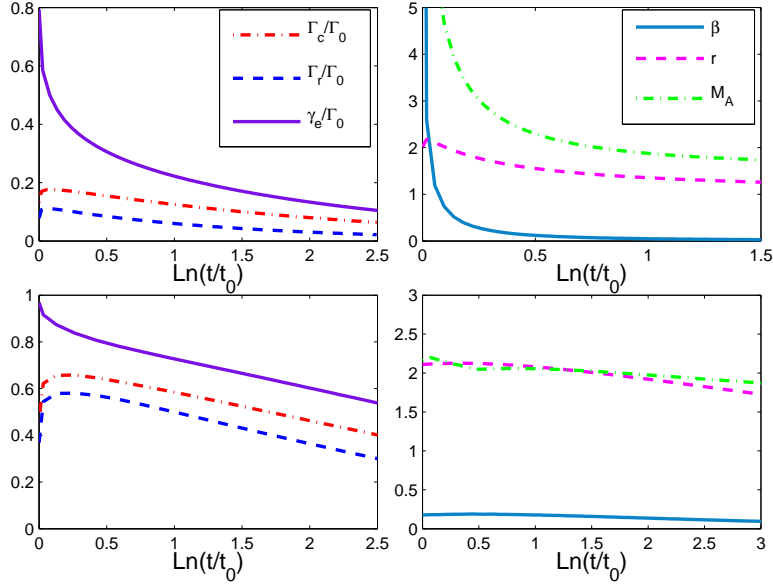}
\caption{\label{fig:mag_1} Left panels: time profiles of the Lorentz factors of the unshocked ejecta $\gamma_e$,
contact discontinuity $\Gamma_c$ and the reverse shock $\Gamma_r$, for adiabatic index $\hat{\gamma}=5/3$, 
$k=0$ and $3\rho_{0}/(8\sigma_0^3a_{i})=0.1$ (upper window), $3\rho_{0}/(8\sigma_0^3a_{i})=10^3$ (lower window).     
Right panels: the evolution of the shock compression ratio $r$ (see 
Eq. [\ref{sh-compression}]), the ratio of kinetic-to-magnetic pressure $\beta=2p_2/b_2^2$, and the fast magnetosonic Mach number $M_A$ of the
fluid upstream the reverse shock (see Eq. [\ref{Alfven-M}]).}
\end{figure}
 
\begin{equation}
\sigma_e(\chi)=\frac{(\chi^{1/3}-1)^2}{4\chi^{1/3}},\quad \rho_e(\chi)=\rho_0\frac{\sigma_e(\chi)}{\sigma_0}(t/r_0)^{-2}.
\quad \gamma_e=2\sigma_0\chi^{-1/3}.\label{impulsiv_self_shell}
\end{equation}
This solution is applicable above the front of the rarefaction wave, at $\chi<\chi_\star(t)$, where 
\begin{equation}
\chi_\star(t)=[1+(8\sigma_0^2r_0/t)^{2/3}]^{3/2}\simeq 8\sigma^2_0 (t/r_0)^{-1}
\end{equation}
(see Eq. [\ref{chi_star_app}]).  The head of the shell is located at $\chi=1$.  

\begin{figure}[ht]
\centering
\includegraphics[width=12cm]{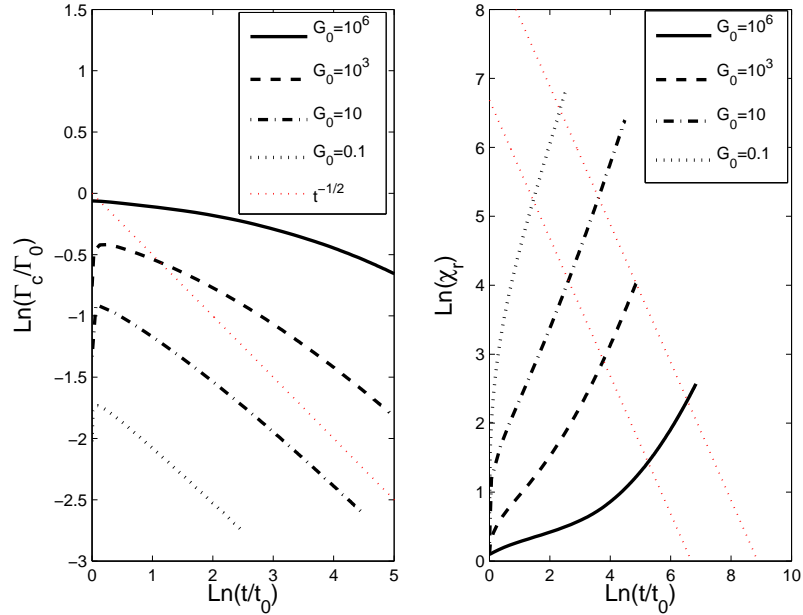}
\caption{\label{fig:mag_2} Left panel: time profiles of $\Gamma_c$ for $\hat{\gamma}=5/3$, $k=0$, and 
different values of $G_0$, where $G_0=3\rho_{0}/(8\sigma_0^3a_{i})$.  Right panel: the corresponding 
trajectories of the reverse shock $\chi_r(t)$.  The red dotted lines mark the location of the rear boundary of the shell (the front
of the reflected rarefaction wave) $\chi_\star(t)$,  for $\sigma_0=10$ (lower curve) and $\sigma_0=30$ (upper curve). The intersection of  the two trajectories $\chi_r(t)$ and $\chi_\star(t)$ gives the time at which crossing of the  accelerating shell by the reverse shock has completed.} 
\end{figure}

To study the effect of the ambient medium on the evolution of the shell during the acceleration phase we assume that the structure of the unshocked shell is given by  (\ref{impulsiv_self_shell}).    We then obtain 
\begin{equation}
G(\chi_r,t)=\frac{3\rho_0}{8\sigma_0^3a_i}\chi_r^{2/3}\sigma_e(\chi_r)(t/r_0)^{k-2}.
\end{equation}
The trajectory of the reverse shock is governed by the equation (see Eq. [\ref{dchi}])
\begin{equation}
\frac{d}{dt}(t\chi_{r})=\frac{q_e(\chi_r)\Gamma_0^2}{\gamma^2_e(\chi_r,t)}
=\frac{q_e(\chi_r)}{2}\chi_r^{2/3}.\label{dchimag}
\end{equation}
At $t=t_0$, $\chi_r=1$ and $\rho_e=\sigma_e=0$.  A strong shock forms initially at the shell's head and quickly propagates 
inwards (in the fluid rest frame).   This brief initial phase is an artifact of out initial conditions
that implicitly assume that impact with the external medium starts only after reflection of the rarefaction wave by the central source. 
As the density and magnetization increases the shock weakens and the contact discontinuity accelerates until 
reaching a terminal value  (Figs \ref{fig:mag_1} and \ref{fig:mag_2}).   This phase lasts for a very short time.  Subsequently,
the contact discontinuity starts decelerating if $k<2$ or maintain a constant speed if $k=2$. For $k>2$ the shell will eventually approach 
free expansion with $\Gamma_c\simeq2\sigma_0$.   For $q_e>>1$, $q_2>>1$ we have from Eq. (\ref{hq}) $h^2\simeq q_e/q_2$.  
Equation (\ref{contac-p}) with $f=0$ then admits the solution
\begin{equation}
\Gamma_c(t)=\left(\frac{3\rho_0\sigma_0}{8a_i}\right)^{1/4}(t/r_0)^{(k-2)/4},\label{Gamc_anl}
\end{equation}
which is viable as long as $\Gamma_c<\Gamma_0=2\sigma_0$.  At early times, when $q_e$ and $q_2$ are of order unity the solution must be
obtained numerically.  Note that for $k=2$ Eq. (\ref{Gamc_anl}) reduces to Eq. (47) in Lyutikov (2010a), that gives the Lorentz 
factor of a one-dimensional planar shell expanding in a constant density medium.  This is expected since for the case
considered here (a cold shell with a purely toroidal magnetic field) 
the flow equations in spherical geometry, (\ref{conti})-(\ref{state}), reduce to those in planar geometry upon the change of 
variables: $\rho\rightarrow r^2\rho$, $b\rightarrow rb$.

Fig \ref{fig:mag_2} delineates solutions for $k=0$ (a constant density medium) and different values of $G_0\equiv3\rho_0/(8\sigma_0^3a_i)$.  
For $G_0>>1$ the effect of the ambient medium is initially small; the reverse shock is confined at the head of the shell, viz., 
$\chi_r\sim 1$ (see right panel of Fig \ref{fig:mag_2}), and $\Gamma_c\sim\Gamma_0$ and declines slowly.  
For sufficiently high $\sigma_0$ the reverse shock will eventually start accelerating inwards in the shell frame before crossing 
of the shell has completed, so that $\gamma_e$ becomes significantly larger than $\Gamma_r$ and $q_e>>1$, $q_2>>1$.
As seen in the left panel of Fig. \ref{fig:mag_2}, the evolution of the contact discontinuity
then approaches $\Gamma_c\propto t^{-1/2}$, as expected from Eq. (\ref{Gamc_anl}).      
This transition occurs at a time $t=t_{dec}$, where
\begin{equation}
t_{dec}/r_0=\left(\frac{3\rho_0}{8\sigma_0^3a_i}\right)^{1/(2-k)}=\left(\frac{9E}{32\pi r_0^3\sigma_0^4a_i}\right)^{1/(2-k)},
\end{equation}
at which $G(R_r,t_{dec})$ in Eq. (\ref{contac-p}) approaches unity. Here $E=\rho_0\sigma_04\pi r_0^3/3$ 
is the total energy of the shell.  The transition will occur before complete shock crossing, that is at $1<\chi(t_{dec})<\chi_\star(t_{dec})$, 
if $t_{dec}<8\sigma_0^2r_0$, or
\begin{equation}
\sigma_0>\sigma_{crt}=\frac{1}{2}\left(\frac{9E}{32\pi r_0^3a_i}\right)^{1/(8-2k)}.\label{sig_crt_k}
\end{equation}
\begin{figure}[ht]
\centering
\includegraphics[width=11cm]{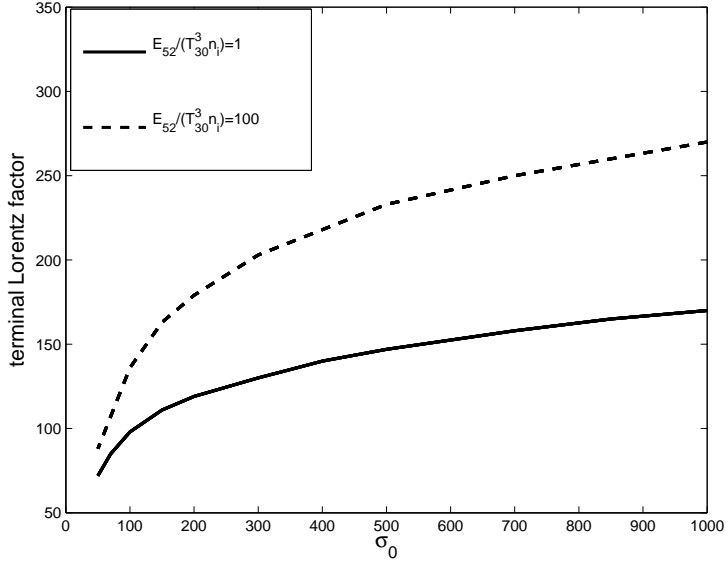}
\caption{\label{fig:terminal}The dependence of the terminal Lorentz factor of the accelerating shell $\Gamma_{max}$, 
measured at time of complete shock crossing, on the initial magnetization $\sigma_0$, for $k=0$, $\hat{\gamma}=5/3$ 
and $E_{52}/(T^3_{30}n_i)=1$ (solid line), $E_{52}/(T^3_{30}n_i)=100$ (dashed line).} 
\end{figure}
The latter scaling can be derived also from energy considerations.  The energy accumulated inside the shocked ambient medium layer
(i.e., between the forward shock and the contact discontinuity) at time $t$ is 
approximately $E_f=4\pi\Gamma_0^2a_ir_0^3(t/r_0)^{3-k}/(3-k)$.  Deceleration of the shocked layer will commence before the shell 
reaches the coasting phase if $E_f\simeq E$ at time $t<8\sigma^2_0r_0$.  With $\Gamma_0=2\sigma_0$ the latter condition yields
\begin{equation}
\sigma_0>\frac{1}{2}\left[\frac{(3-k)E}{32\pi r_0^3a_i}\right]^{1/(8-2k)},
\end{equation}
in rough agreement with (\ref{sig_crt_k}). 
For a burst of duration $T=30 T_{30}$ sec and total energy $E=10^{52}E_{52}$ ergs, expanding in an ambient medium
of constant number density $n_i$ measured in c.g.s units the condition (\ref{sig_crt_k}) reduces to
\begin{equation}
\sigma_0>\sigma_{crt}=90\left(\frac{E_{52}}{T_{30}^3n_i}\right)^{1/8},\label{sig_crt}
\end{equation}
where the initial shell width has been taken to be $r_0=cT$.  
Consequently, sub-critical shells (i.e., $\sigma_0<\sigma_{crt}$) will not be significantly affected by the ambient medium before
reaching the coasting phase.  High-sigma shells ($\sigma_0>>\sigma_{crt}$), on the other hand, will experience 
significant deceleration of the head well before the onset of the coasting phase, and it is anticipated
that the  Lorentz factor of the shell will be limited to $\Gamma_{\rm max}\simeq2\sigma_{crt}$.  The dependence of 
the terminal Lorentz factor on the initial magnetization $\sigma_0$, obtained from numerical integrations of   
Eqs. (\ref{Psi0}), (\ref{contac-p}) and (\ref{dchimag}), is exhibited in Fig. (\ref{fig:terminal}).  As seen, for high-sigma shells 
the scaling $\Gamma_{\rm max}\simeq2\sigma_{crt}=180(E_{52}/T_{30}^3 n_i)^{1/8}$, derived above using heuristic arguments, 
is in quite good agreement with the numerical result.  Note that the critical magnetization (\ref{sig_crt}) depends on the
initial energy density of the shell, $E/r_0^3$.  Thus, for a given power $L=E/T=cE/r_0$ we have 
$\sigma_{crt}\propto L^{1/8}r_0^{-1/4}$, implying a less restrictive constraint on the Lorentz factor for smaller (sub) shells.
For all cases studied it is found that a reverse shock always exists 
(i.e., $M_A>1$) in the acceleration phase, however, except for the very early stages of the evolution the shock is weak and
magnetically dominated, that is $\beta=2p_2/b_2^2<<1$.  Emission from the shocked ejecta is not anticipated in high sigma 
shells, at least not in the ideal MHD case.  

\begin{figure}[ht]
\centering
\includegraphics[width=12cm]{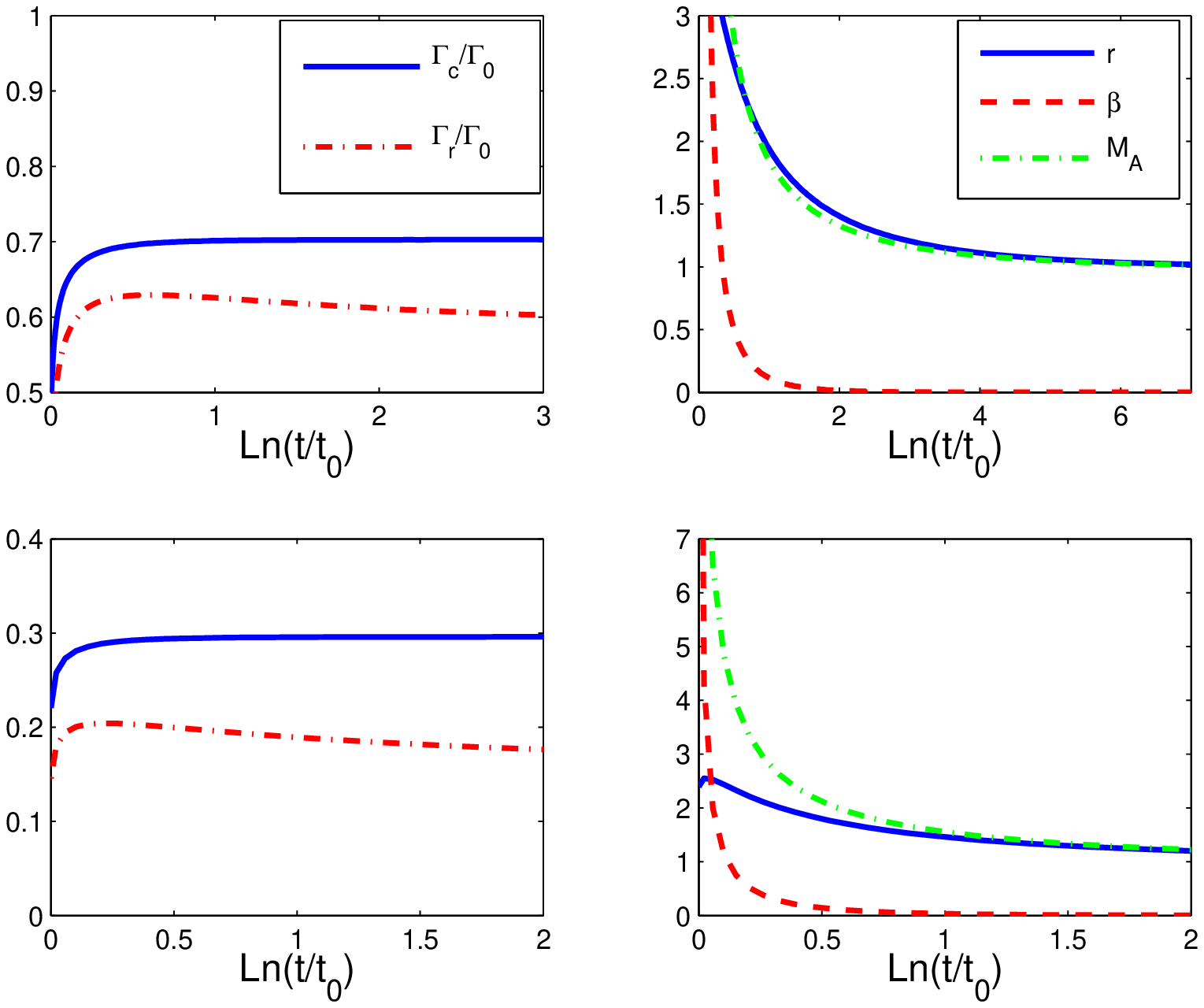}
\caption{\label{fig:mag_k}
 Left panels: time profiles of the Lorentz factors of the
contact discontinuity $\Gamma_c$ and the reverse shock $\Gamma_r$, for adiabatic index $\hat{\gamma}=5/3$, 
$k=2$ and $3\rho_{0}/(8\sigma_0^3a_{i})=10^3$ (upper window), $3\rho_{0}/(8\sigma_0^3a_{i})=1$ (lower window).     
Right panels: the evolution of the shock compression ratio $r$, the ratio of kinetic-to-magnetic pressure 
$\beta=2p_2/b_2^2$, and the Alfven Mach number $M_A$.}
\end{figure}

In case of a stellar wind ($k=2$) the shocked layer, after a brief rearrangement phase, maintains a constant
speed (Fig. \ref{fig:mag_k}), as expected from Eq. (\ref{Gamc_anl}).  The reverse shock in this case quickly weakens
and eventually dies away ($M_A$ becomes smaller than unity).  The communication with the contact discontinuity 
then proceeds via a magnetosonic wave.  For a mass loss rate $\dot{M}_w=\dot{M}_{-5}10^{-5}$  M$_{\sun}$ yr$^{-1}$ 
and terminal velocity $v_w=10^8v_{w8}$ cm s$^{-1}$ the ambient density scales as 
$n_i=10^{11}\dot{M}_{-5}(r/R_0)^{-2}v_{w8}^{-1}$ cm$^{-3}$, with $R_0=10^{13}R_{13}$ cm being the radius at the wind's base.
With this parametrization Eq. (\ref{Gamc_anl}) yields for the Lorentz factor of the contact
\begin{equation}
\Gamma_c\simeq10\left(\frac{E_{52}v_{w8}}{T_{30}R_{13}^2\dot{M}_{-5}}\right)^{1/4}.\label{Gamc_wind}
\end{equation}

\begin{figure}[ht]
\centering
\includegraphics[width=12cm]{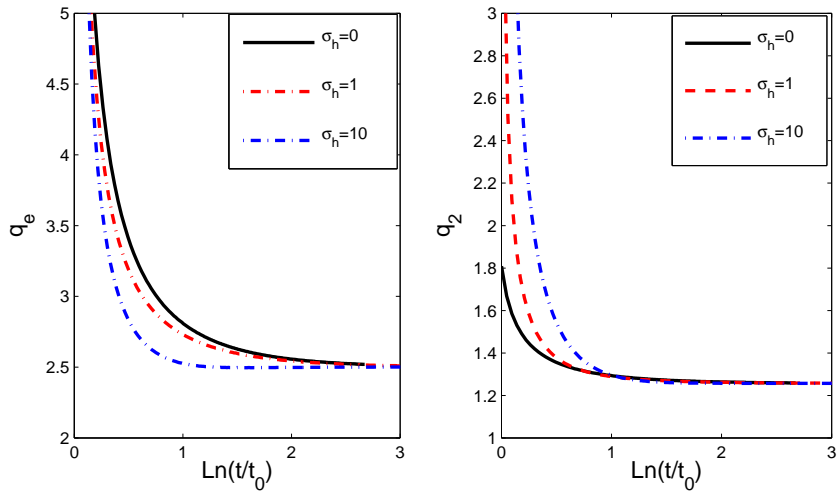}
\caption{\label{fig:free}The evolution of $q_e=(\gamma_e/\Gamma_r)^2$ (left panel) and $q_2=(\Gamma_c/\Gamma_r)^2$ (right panel)
for adiabatic index $\hat{\gamma}=5/3$, $k=0$, $n=2$, $3a_e/4a_i\Gamma_0^2=0.1$, and different magnetizations.} 
\end{figure}

\begin{figure}[ht]
\centering
\includegraphics[width=12cm]{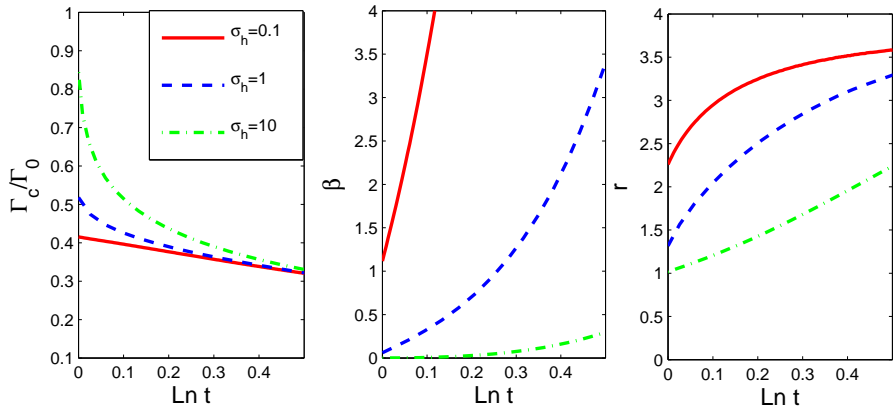}
\caption{\label{fig:free2}Evolution of the Lorentz factor of contact discontinuity (left), ratio of kinetic-to-magnetic pressures
$\beta=2p_2/b_2^2$ (middle), and shock compression ration r (right), for a shell of initial width $\Delta \chi=10$, $k=0$, $n=2$,
and different values of the magnetization $\sigma_h$.  The time at which reverse shock crossing is completed is $\ln (t/t_{dec})=0.5$ }
\end{figure}

As explained above, in a uniform circumburst medium a shell with initial magnetization $\sigma_0<\sigma_{crt}$ 
will start decelerating only after reaching 
the coasting phase, when conversion of magnetic energy into bulk kinetic energy has nearly completed.  Once approaching 
the coasting phase the shell starts spreading and its structure is altered (GKS10).  To compute the evolution of the 
system in this region we adopt the self-similar solution derived in GKS10 (see also appendix \ref{sec:app-self-eject}) for the unshocked, 
coasting ejecta.  As noted in  appendix \ref{sec:app-self-eject}, this solution
is not fully self-consistent in vacuum, as it implicitly assumes a confining agent at the head.  However, is can be matched self-consistently
to a shocked layer through the jump conditions at the reverse shock, and so may provide a reasonable description for the 
interacting ejecta after its spreading.  With this choice $\gamma_e=\Gamma_0\chi^{-1/2}$,  $G(\chi,t)$ is given by Eq. (\ref{G-self}) 
(see appendix \ref{sec:app-self-eject}), and
\begin{equation}
\sigma_e(\chi,t)=\sigma_h\chi^{-(n/2+1)}(t/t_{dec})^{-1},\label{sig_coast}
\end{equation}
where $\sigma_h$ denotes the magnetization at the head ($\chi=1$) of the freely coasting shell at time of impact, $t=t_{dec}$.
The energy density of the unshocked shell can be expressed as
\begin{equation}
T^{00}=\rho_e(1+\sigma_e)\gamma_e^2=a_e\Gamma_0^2(t/t_{dec})^{-3}\chi^{(n/2-1)}(1+\sigma_e),\label{T00_un_free}
\end{equation}
with $\sigma_e$ given by (\ref{sig_coast}), so that for $\sigma_e<<1$ the choice $n=2$ corresponds roughly to a uniform 
energy distribution (since the shell is thin).  Inspection of Eq. (\ref{sig_coast}) reveals that for $n>-2$ the magnetization
of the shell decreases with increasing $\chi$.  Consequently, it is anticipated that the solution will eventually converge to that 
of NS06 discussed in \S\S \ref{sec:unmag}.  This is confirmed in Fig. \ref{fig:free}.  In practice the shell has a finite width $\Delta$.
At the onset of the coasting phase the width of the shell is still $\Delta_c\sim 2r_0$, corresponding to $\Delta \chi\sim(2r_0/r_c)2\Gamma^2_0
\simeq1$, using $r_c\sim\sigma^2_0$ and  $\Gamma_0\sim\sigma_0$ for the coasting radius and terminal Lorentz factor 
(GKS10 and appendix \ref{sec:app-self-eject}).  At this point the width of the shell 
starts growing and at the deceleration radius $r_{dec}$ it is expected to be $\Delta\sim\eta\Delta_c$, corresponding 
to  $\Delta \chi\sim\eta$, where $\eta$ is a fraction of the ratio $r_{dec}/r_c$.  
Fig \ref{fig:free2} exhibits solutions for a shell of width $\Delta\chi=10$ interacting with a constant density medium ($k=0$).
Although $\sigma_e<1$ is expected in the coasting phase, we present also a case with $\sigma_h=10$ to elucidate the 
general behavior of the system.  As seen, for $\sigma_h<1$ the compression ratio of the reverse shock increases 
considerably and the pressure downstream becomes
kinetic dominated before complete crossing of the reverse shock, at $\ln (t/t_{dec})=0.5$.  We find this behavior to be quite robust for 
self-similar ejecta with $n\ge0$.

\subsection{Multiple shell model}
The rapid variability observed in many GRBs is commonly attributed to ejection of many sub-shells of small width that collide at relatively 
large radii.   Such a multi-shell scenario can be envisaged also for magnetically dominated  ejecta.  In that case collisions of
shells should occur only after full conversion of magnetic energy into bulk kinetic energy is accomplished, and before the onset of 
deceleration by the ambient medium.   As explained above
the restriction on the dynamics of a single shell imposed by the surrounding  medium  (Eq. [\ref{sig_crt}]) is less stringent 
for a shell of small width, so that in principle it is anticipated that for sufficiently rapid ejections the resultant shells may accelerate 
to the maximum  Lorentz factor, $\Gamma\sim \sigma_0$ ( at which the initial magnetic energy is fully converted),
before a substantial  fraction of their energy is dissipated at the forward shock.   However, in order that the
shells will collide only after reaching the coasting phase, the time interval between successive ejections of shells must be
large enough, otherwise the shells will merge to form a steepening magnetosonic wave train 
with a characteristic size of the order of the engine's life time.

To estimate the duty cycle required for shells to collide at the coasting phase, consider two  shells of initial width $l_i$ 
and magnetization $\sigma_{i0}$ ($i=1,2$) expelled into vacuum.   The shells will collide provided $\sigma_{20}>\sigma_{10}$.  
Suppose  that the first shell is ejected at time $t_0$ and the second one at $t_0+\Delta t$.     Now, the rarefaction wave
of the first shell accelerates as $\gamma_{w1}=\sqrt{2}\sigma_{10}^{2/3}( t/l_1)^{1/6}$ (see Eq. [\ref{app-g_w}]).    
Let $\gamma_{e2}(\chi_2)=2\sigma_{20}\chi_2^{-1/3}$ denotes the Lorentz factor
of some point (that is, a fixed value of the self-similarity parameter $\chi_2$) above the front of the rarefaction wave of the second shell.    
The relative velocity between that point  and the rarefaction front of the first shell is 
\begin{equation}
v_{e2}-v_{w1}\simeq \frac {1}{2\gamma_{w1}^2}-\frac{1}{2\gamma_{e2}^2}=
\frac{1}{4}\sigma_{10}^{-4/3} (t/l_1)^{-1/3}
\left[ 1-\frac{\sigma_{10}^{4/3}}{\gamma_{e2}^2}\left(\frac{t}{l_1}\right)^{1/3}\right].
\end{equation}
The two parts will catch up at time
\begin{equation}
t_{coll}=\int_{0}^{\Delta t}\frac{dt^\prime}{v_{e2}-v_{w1}}\simeq 4l_1\sigma_{10}^{4/3}
\int_0^{y_1}\frac{y^{1/3}}{1-(2\sigma_{10}^{4/3}/\gamma_{e2}^2)y^{1/3}}dy,
\end{equation}
where $y_1=\Delta t/l_1$.  For $2\sigma_{10}^{4/3}y_1^{1/3} <<\gamma_{e2}^2$ the result 
is approximately 
\begin{equation}
t_{coll}\simeq 3l_1(\sigma_{10}\Delta t/l_1)^{4/3}.
\end{equation}
The coasting time (radius) of the first shell is $t_{cost}\simeq 8\sigma_{10}^2l_1$.  Thus, collision will occur after the first shell has reached 
the coasting phase provided $t_{coll}>t_{cost}$ or
\begin{equation}
(\Delta t/l_1)>(8/3)^{3/4}\sigma_{10}^{1/2}.\label{duty-cyc}
\end{equation}
For $\sigma_{10}>>1$  Eq. (\ref{duty-cyc}) implies an unlikely small duty cycle.    The above discussion suggests that within the framework 
of ideal MHD intermittent ejection 
of magnetically dominated outflow leads ultimately to one impulsive shell of size roughly equals the life time of the system.

\section{\label{sec:conc}Conclusions}
We considered the interaction of a relativistic magnetized shell with an ambient medium,  focusing on
the case of an impulsive, high-sigma shell.  We find
that for values of the initial magnetization $\sigma_0$ larger than the critical value given by Eq. (\ref{sig_crt_k}) 
the evolution of the system is  significantly altered by the ambient medium  well before the shell reaches its coasting phase.  
For such high sigma shells a major fraction of the explosion energy is dissipated behind the forward shock by the time 
compression of the shell by the ambient medium is communicated to the accelerating rarefaction wave.  The maximum 
Lorentz factor of the shell is then limited to values well below  $\sigma_0$ (see  Fig. \ref{fig:terminal}).    
Such episodes are expected to produce a smooth, relatively fast rising slowly decaying (power law) light curve, even in a multi-shell scenario.
Events like GRB080916C and GRB090510 (Abdo et al. 2009a,b) are not easily accounted for by the impulsive high-sigma shell model.
If extracted magnetically, such outflows may require magnetic dissipation beyond the ideal MHD limit, as may 
occur in e.g., a striped wind model.  Intermittent ejection doesn't seem to help, as unlikely small duty cycle is required in order 
for shells to collide after reaching the coasting phase.

Lower sigma shells start decelerating only after reaching the coasting phase.   The properties of the reverse shock then depend on the 
structure of the coasting shell.  For a self-similar shell with a reasonable density profile the magnetization decreases inwards and it is anticipated that the reverse shock will become strong before complete crossing.  The evolution of the shocked layers is shown to quickly approach (for shells with $\sigma <1$) the self-similar solution derived by Nakamura and Shigeyama (2006).  The slow acceleration of high sigma shells, $(\Gamma\propto r^{1/3})$, relative to 
a radiatively driven outflow ($\Gamma\propto r$), implies a smaller optical depth at the radius of collision.  This can alleviate the need for
extremely high Lorentz factors for most GRBs, to avoid strong absorption of the highly variable prompt emission.

I thank Yoni Granot, Yuri Lyubarsky, Maxim Lyutikov, and Udi Nakar for useful comments, and the anonymous referee for 
a detailed report. 
This work was supported by an ISF grant for the Israeli Center for High Energy Astrophysics.

\appendix

\section{\label{sec:app-jump-r}Jump conditions at the reverse shock}
The stress-energy tensor of a magnetized fluid takes the form 
\begin{equation}
T^{\mu\nu}=\rho \tilde{h}u^\mu u^\nu-g^{\mu\nu}\tilde{p}-b^\mu b^\nu,\label{Tmunu}
\end{equation}
where $\rho$ is the proper density, $\tilde{p}=p+p_b$ is the sum of kinetic pressure $p$ and magnetic pressure $p_b=b^\mu b_\mu/2=b^2/2$,  
$\tilde{h}=e+p/\rho+b^2/\rho$ is the generalized specific enthalpy, $u^\alpha$ is the fluid 4-velocity and $g^{\mu\nu}$ is the metric tensor. 
The 4-vector of the magnetic field is defined in terms of the electromagnetic tensor $F^{\mu\nu}$ as 
\begin{equation}
b_\alpha=\frac{1}{2}\eta_{\alpha\beta\gamma\delta}u^\beta F^{\gamma\delta},
\end{equation}
where $\eta_{\alpha\beta\gamma\delta}$ is the Levi -Civita tensor.  The jump conditions are obtained 
from integration of the flow equations,
\begin{eqnarray}
\partial_\mu(\rho u^{\mu})=0,\label{cont}\\
\partial_\mu T^{\mu\nu}=0\label{enrg-mom},\\
\partial_\mu(b^\mu u^{\nu}-b^\nu u^\mu)=0,\label{faraday}
\end{eqnarray}
across the shock surface  $\psi(x^\mu)\equiv r-R(t,\theta,\phi)=0$, whereby we have
\begin{eqnarray}
[\rho u^{\mu}]n_\mu=0,\quad \left[b^\mu u^\nu -b^\nu u^{\mu}\right]n_\nu=0,\quad
\left[T^{\mu\nu}\right]n_\nu=0.\label{bcT}
\end{eqnarray}
The square brackets denote the difference of the enclosed quantity across the shock front, and
\begin{equation}
n^\mu=\frac{\partial_\mu\psi}{\sqrt{\partial_\mu\psi\partial^\mu\psi}}\label{normal}
\end{equation}
is a 4-vector normal to the shock front.  For the reverse shock with $R(t,\theta,\phi)=R_r(t)$ 
\begin{equation}
n_\mu=\Gamma_r(-V_r,1,0,0),\label{normal_r}
\end{equation}
here $V_r=dR_r/dt$ is the velocity of the reverse shock and $\Gamma_r=(1-V_r^2)^{-1/2}$.
Equations  (\ref{Tmunu}),  (\ref{bcT}) and (\ref{normal_r}) yield for the quantities defined in Fig. \ref{fig:a}
\begin{eqnarray}
b_e\gamma_e(v_e-V_r)=b_2\gamma_2(v_2-V_r),\label{jmpB-0}\\
\rho_e\gamma_e(v_e-V_r)=\rho_2\gamma_2(v_2-V_r),\label{jmpB-1}\\
\rho_e\tilde{h}_e\gamma_e^2v_e(v_e-V_r)+\tilde{p}_e=\rho_2\tilde{h}_2\gamma_2^2v_2(v_2-V_r)+\tilde{p}_2,\label{jmpB-2}\\
\rho_e\tilde{h}_e\gamma^2_e(v_e-V_r)+\tilde{p}_eV_r=\rho_2\tilde{h}_2\gamma^2_2(v_2-V_r)+\tilde{p}_2V_s.\label{jmpB-3}
\end{eqnarray}

We assume a cold ejecta, $p_e=0$, and adopt $h_2=\rho_2+ap_2$ where $a=\hat{\gamma}/(\hat{\gamma}-1)$ and $\hat{\gamma}$  
is the adiabatic index, for the shocked ejecta  (see Fig \ref{fig:a}).
Then, to order $O(\Gamma_r^{-2})$  the solution of  Eqs. (\ref{jmpB-0})-(\ref{jmpB-3}) is given by
\begin{eqnarray}
\Psi(q_2,q_e,\sigma_e)=0,\label{Psi}\\
\rho_{2}=\rho_e h(q_2,q_e),\\
b_2=b_e(\rho_2/\rho_e)=b_e h(q_2,q_e),\\
p_{2}=\rho_ef(q_2,q_e,\sigma_e),\label{press2}
\end{eqnarray}
with
\begin{eqnarray}
h(q_2,q_e)=\frac{\sqrt{q_2}}{\sqrt{q_e}}\left(\frac{q_e-1}{q_2-1}\right),\label{hq}\\
f(q_2,q_e,\sigma_e)=\frac{(q_e-1) }{a(q_2-1)+2}\left[1-\sqrt{q_2/q_e}+\frac{q_e\sigma_e}{q_e-1}
-\frac{\sigma_eq_2^2(q_e-1)}{q_e(q_2-1)^2}\right],\label{fq}\\
\Psi(q_2,q_e,\sigma_e)=2q_e(1+\sigma_e-\sqrt{q_2}/\sqrt{q_e})+(a-2)q_2\sigma_e\left(\frac{q_e-1}{q_2-1}\right)\cr
+(1+\sigma_e)[a(q_2-1)+2]\left(\frac{q_2-q_e}{q_2+1}\right)
-a\sigma_eq_e\left(\frac{q_2-1}{q_e-1}\right),\label{psiq}
\end{eqnarray}
where $q_2=(\Gamma_c/\Gamma_r)^2$, $q_e=(\gamma_e/\Gamma_r)^2$,  and $\sigma_e=b_e^2/\rho_e$.  The relations
(\ref{hq})-(\ref{psiq}) generalize those derived by Kennel \& Coroniti (1984) for a high Alfven Mach number 
shock, corresponding to $q_e>>q_2$ in our notation.
In this limit Eqs. (\ref{Psi}) and (\ref{psiq}) with $\hat{\gamma}=4/3$ reduce to 
\begin{equation}
\frac{2-q_2}{q_2}+\left(\frac{q_2+1}{q_2-1}\right)\left(\frac{\sigma_e}{1+\sigma_e}\right)=0.
\end{equation}
This condition can be expressed in terms of the 4-velocity of the downstream fluid, measured with respect to the shock frame, 
$u_2^\prime=\Gamma_r\gamma_2(v_2-V_r)=(q_2-1)/2\sqrt{q_2}+O(\Gamma_r^{-2})$, as 
\begin{equation}
1+4u_2^{\prime 2}-\frac{\sqrt{1+u_2^{\prime 2}}}{u_2^{\prime }}\left(4u_2^{\prime 2}-\frac{\sigma_e}{1+\sigma_e} \right)=0,
\end{equation}
which is equivalent to Eq. (4.10) in Kennel \& Coroniti (1984).   The solution for $u_2^{\prime }$ is given by Eq. (4.11) in the same reference.

\section{\label{sec:app-self-eject}Self-similar ejecta}
We consider a spherically symmetric,  magnetized ejecta with a purely toroidal 
magnetic field, viz., $b_e^\mu=(0,0,0,b_e)$.   The flow equations (\ref{cont})-(\ref{faraday}) then reduce to 
\begin{eqnarray}
\frac{\partial (\rho_e\gamma_e)}{\partial t} +\frac{1}{r^2}\frac{\partial}{\partial r} (r^2\rho_e\gamma_e v_e)=0,\label{conti}\\
\frac{\partial (b_e\gamma_e)}{\partial t} +\frac{1}{r}\frac{\partial}{\partial r} (rb_e\gamma_e v_e)=0,\label{magB}\\
\rho\tilde{h}_e\gamma_e^2\frac{d v_e}{dt}+v_e\frac{\partial \tilde{p_e}}{\partial t} +\frac{\partial 
\tilde{p}_e}{\partial r} +\frac{b_e^2}{r}=0,\label{momentum}\\
\frac{d}{dt}\ln\left(p_e/\rho_e^{\hat{\gamma}}\right)=0.\label{state}
\end{eqnarray}
We seek self-similar solutions that are separable, to order O($\Gamma^{-2}$, in the variables $\tau=\ln(t)$ and 
\begin{equation}
\chi=\{1+2(m+1)\Gamma_{0}^2\}[1-(r-r_a)/t]\label{chi},
\end{equation}
with the front of the expanding ejecta located at $\chi=1$, and $\Gamma_0=\gamma_e(\chi=1)=At^{-m/2}=Ae^{-m\tau/2}$.
We adopt the following parametrization of the fluid variables:
\begin{eqnarray}
\gamma_e^2=\Gamma_0^2 g(\chi),\label{g}\\
b_e=b_0 e^{-p\tau}B(\chi),\label{B}\\
\rho_e^\prime=\rho_e\gamma_e=\rho^\prime_oe^{-q\tau}H(\chi).\label{h}
\end{eqnarray}
Transforming from the coordinates $(r,t)$ to $(\chi,\tau)$  and using the relations
\begin{eqnarray}
t\partial_t=\partial_\tau+[(m+1)(2\Gamma_0^2-\chi)+1]\partial_\chi,\label{dt2}\\
t\partial_r=-[1+2(m+1)\Gamma_0^2]\partial_\chi,\label{dr2}\\
\end{eqnarray}
one obtains, upon substitution of Eqs. (\ref{g})-(\ref{h}) into the flow equations (\ref{conti})-(\ref{state}),
\begin{eqnarray}
\frac{A}{g}\frac{d\ln g}{d\chi}= m(1-g\chi)-4\sigma_e(1-p-m/2),\label{out1}\\
\frac{A}{g}\frac{d\ln B}{d\chi}=m-(1-p)(1-g\chi)-2\sigma_e(1-p-m/2),\label{out2}\\
\frac{A}{g}\frac{d\ln H}{d\chi}=m-(2-q)(1-g\chi)+\frac{4\sigma_e}{(1-g\chi)}[m/2+p-1+(2-q)g\chi].\label{out3}
\end{eqnarray}
to order $O(\Gamma_0^{-2})$, where $A=(m+1)[(1-g\chi)^2-4\sigma_eg\chi]$.

\subsection{Freely expanding ejecta}
For a freely expanding ejecta $d\gamma_e/dt=-m/2+(m+1)(1/g-\chi)\partial_\chi\sqrt{g}=0$.  The boundary condition 
$g(\chi=1)=1$ implies $m=0$ and $g(\chi)=\chi^{-1}$.   Equations (\ref{out1})-(\ref{out3})  give $p=2$, $q=3$, $B(\chi)=\chi^{-1/2}$.  
The function  $H(\chi)$ is unrestricted.   To summarize, the solution can be expressed as
\begin{eqnarray}
\gamma_e^2=\Gamma_0^2\chi^{-1},\quad b_e=b_0 e^{-2\tau}\chi^{-1/2},\quad \rho_e^\prime=a_e\Gamma_0e^{-3\tau}H(\chi),\quad
\sigma_e=\sigma_0e^{-\tau}\chi^{-3/2}H^{-1}(\chi),
\end{eqnarray}
with  $b_0$ denoting the value of $b_e$ at $\tau=0$ $\chi=1$.  The discussion in NS06 suggests that a power law density profile, $\rho_e\propto\gamma_e^{-n}$, corresponding to $H=\chi^{(n-1)/2}$, provides a reasonable description of realistic ejecta.   Note that magnetic and kinetic
energies have different scaling,  $\rho_e\gamma_e^2\propto t^{-3}$ and $b_e^2\gamma_e^2\propto t^{-4}$, so that the enthalpy is not self-similar.
Note also that magnetic energy is not conserved, viz., $b_e^2 t^3\propto t^{-1}$, implying a loss of Poynting energy from the front.   This is
a consequence of an implicit boundary condition at the vacuum-shell interface that assumes a confining agent.  The loss of magnetic energy 
is then associated with a $pdV$ work.  For the interacting shell this solution can be matched with the shocked shell layer, as discussed 
in \S \ref{sec:mag-shell}.

\subsection{\label{sec:app_impulsiv}An accelerating high-$\sigma$ shell}
The solution describing an impulsive accelerating shell (GKS10, Lyutikov 2010a) corresponds to the choice 
$m=0$, $p=1$, $q=2$.  Eqs. (\ref{out1})-(\ref{out3}) then readily yield
\begin{equation}
\sigma_e(\chi,\tau)=\frac{(\chi^{1/3}-1)^2}{4\chi^{1/3}},
\end{equation}
independent of $\tau$, and
\begin{eqnarray}
g(\chi)=\chi^{-2/3},\\
B(\chi)=H/\sqrt{g}=\sigma_e(\chi).
\end{eqnarray}
The requirement $\gamma_e=\Gamma_0\chi^{-1/3}>1$  formally implies  $\chi<\chi_0=\Gamma_0^{1/3}$ and $\Gamma_0<4\sigma_0$.  However, the
above solution is applicable only well above the magnetosonic point, defined by the condition $\sigma_e=\gamma_e^2v_e^2\simeq\gamma_e^2$.
A full treatment (GKS10) gives $\Gamma_0=2\sigma_0$ where  $\sigma_0$ is the initial magnetization
of the shell.    To the order at which we are working the self-similar variable is then given by (\ref{chi_impuls}) and the solution 
by (\ref{impulsiv_self_shell}) with $\rho_0$ denoting the initial density of the shell.

We choose $r_a=r_m=2r_0$ in (\ref{chi}), where $r_0$ is the initial shell width and $r_m$ is the magnetosonic radius.  Then
at $\tau=0$ (corresponding to $t=r_m$) the front of the reflected rarefaction wave is located at  $\chi=8\sigma_0^2$.  The local velocity of the front, as measured in the fluid frame, is $v^\prime_w=\sqrt{\sigma_e/(1+\sigma_e)}$, and the 
corresponding Lorentz factor is $\gamma^\prime_w=(1-v_w^{\prime2})^{-1/2}$.   The Lorentz factor of the wave front in the lab frame is 
obtained upon a Lorentz transformation:
\begin{equation}
\gamma_w=\gamma_e\gamma^\prime_w(1+v_ev^\prime_w)\simeq 2\gamma_e\sqrt{\sigma_e}.\label{app-g_w}
\end{equation}
The trajectory of the wave front is governed by the equation $dr_\star/dt=v_w\simeq1-1/2\gamma_w^2$, which can be translated into
\begin{equation}
\frac{d}{dt}(t\chi_{\star})=(8\sigma_0/\gamma_w)^2=\frac{\chi_\star^{2/3}}{4\sigma_e(\chi_\star)}\simeq\chi_\star^{1/3},\label{wave_traj}
\end{equation}
where the last equality holds at $\chi_\star>>1$ for which $4\sigma_e\simeq \chi_\star^{1/3}$.  The solution of the latter equation reads
\begin{equation}
\chi_\star(t)=[1+(8\sigma_0^2r_0/t)^{2/3}]^{3/2},\label{chi_star_app}
\end{equation}
At $t<<8\sigma_0^2r_0$ we have $\chi_\star\simeq 8\sigma_0^2(r_0/t)$ and  
$\gamma_e(\chi_\star)\simeq (\sigma_0t/2r_0)^{1/3}$, $\gamma_w\simeq\sqrt{2}\sigma_0^{2/3}(t/r_0)^{1/6}$ 
in agreement with GKS10 and Lyutikov (2010a).

The energy density is 
\begin{equation}
T^{00}=\rho_e(1+\sigma_e)\gamma_e^2\simeq\frac{\rho_0\sigma_0}{4}(1-\chi^{-1/3})^4(t/r_0)^{-2}.\label{T00_imp}
\end{equation}

Using $r^2 dr=(t^3/8\sigma_0^2)d\chi$ we obtain for the total energy contained between the reflected rarefaction wave and the head 
\begin{equation}
\Delta E=\int{T^{00}4\pi r^2dr}=\pi\rho_0\sigma_0r_0^2(t/8\sigma_0^2)\int_1^{\chi_\star}{(1-\chi^{-1/3})^4d\chi}
\simeq \pi\rho_0\sigma_0r_0^3=(3/4)E\label{E_imp}
\end{equation}
independent of time, where $E$ is the initial energy (i.e., the explosion energy).

\end{document}